\documentclass[12pt]{article}
\usepackage{fullpage}

\usepackage{graphicx, amsmath, amssymb, psfrag, natbib}

\def\bW{\mathbf{W}}
\def\bw{\mathbf{w}}
\def\bZ{\mathbf{Z}}
\def\bz{\mathbf{z}}

\def\ba{\mathbf{a}}

\def\bgamma{\boldsymbol{\gamma}}

\begin{document}

\title{A Simple Method for Detecting Interactions between a Treatment
and a Large Number of Covariates}

\author{
{\sc Lu Tian}
\thanks{Depts. of Health, Research \& Policy,
94305, lutian@stanford.edu}\\
{\sc Ash A Alizadeh}
\thanks{Dept. of Medicine, Stanford University.
94305, arasha@stanford.edu}\\
{\sc Andrew J Gentles}
\thanks{Integrative Cancer Biology Program, Stanford University.
 andrewg@stanford.edu}\\
and\\
{\sc Robert Tibshirani}\thanks{Depts. of Health, Research \&
  Policy, and Statistics,
    Stanford University, tibs@stanford.edu}
}

\maketitle
\begin{abstract}
We consider a setting in which we have a treatment and a large number of covariates
for a set of observations,
and wish to model their relationship with an outcome of interest.
We propose a simple method for modeling interactions between the
treatment and covariates.
 The idea is to modify the covariate in a simple way, and then fit a standard model
using the modified covariates and no main effects.
We show that coupled with an efficiency augmentation procedure, this method produces valid inferences in a variety of settings.
It can be useful for personalized medicine:  determining  from a large set of biomarkers
the subset of patients that can potentially
benefit from a treatment.
We apply the method to both simulated datasets and gene expression studies of cancer.
The modified data can be used for other purposes, for example
large scale hypothesis testing for determining which of a set of
covariates  interact with a treatment variable.
 \end{abstract}

 \section{Introduction}
\label{sec:intro}
To develop strategies for  personalized medicine, it is important to identify the
treatment and covariate interactions in the setting of randomized clinical
trial \citep{RS:08}. To confirm and quantify the treatment effect is often the primary
objective of a randomized clinical trial. Although important, the final
result (positive or negative) of a randomized trial is a conclusion
with respect to the average treatment effect on the entire study population.
 For example, a treatment  may be no better than the placebo in the overall study population, but it may be better for a subset of patients.
Identifying the treatment and covariate interactions may provide
valuable information for determining this subgroup of patients.

In practice, there are two commonly used approaches to characterize the potential treatment
and covariate interactions. First, a panel of simple patient subgroup analyses,
where the treatment and control arms are compared in different patients
subgroups defined a priori, such as male, female, diabetic and non-diabetic patients, may be performed following the main comparison. Such an exploratory
approach mainly focusses on simple interactions between treatment and one
dichotomized covariate. However it will often suffer from false positive
findings due to multiple testing and will not find complicated
treatment and covariates interaction.

 In a more rigorous analytic approach, the treatment and covariates interactions
can be
examined in a multivariate regression analysis where the product of the
binary treatment indicator and a set of baseline covariates are included
in the regression model. Recent  breakthroughs in biotechnology
 makes a vast amount of data available for exploring for potential
interaction effect with the treatment and assisting in the optimal treatment
selection for individual patients. However, it is very difficult to  detect
the interactions between  treatment and high dimensional covariates via
direct multivariate regression modeling. Appropriate variable selection
methods such as Lasso are needed to reduce the number of
covariates having interaction with the treatment. The presence of main
effect, which often have bigger effect on the outcome than the treatment
interactions,  further compounds the
difficulties in dimension reduction since a subset of variables need
to be selected for modeling the main effect as well.

Recently, \citet{BR04} formalized the subpopulation treatment effect pattern plot (STEPP) for characterizing interactions between the treatment and continuous covariates.  \citet{SRZ:07} proposed a efficient algorithm for multivariate model-building with flexible fractional polynomials interactions (MFPI) and compared the empirical performance of MFPI with STEPP.  \citet{Su:08} proposed the classification and regression tree method
to explore the covariates and treatment interactions in survival analysis. \citet{TT2009}
proposed an efficient algorithm to construct an index score, the sum
of selected dichotomized covariates, to stratify patients population
according to the treatment effect. In a more recent work, \citet{ZZRK:12} proposed a novel approach to directly estimate the optimal treatment selection rule via maximizing the expected clinical utility, which is equivalent to a weighted classification problem.  There are also rich Bayesian literatures for flexible modeling nonlinear and nonadditive/interaction relationship between covariates and responses \citep{Le:95,CGM:98, Gu:00, chen12}. However, most of these existing methods excepting that proposed by \cite{ZZRK:12},
are not designed to deal with high-dimensional covariates.

In this paper,
we propose a simple approach to estimate the covariates and treatment
interactions without the need for modeling main effects.
The idea is simple, and in a sense, obvious. We simply code the treatment
variable as $\pm 1$ and then include  the  products
of this variable with centered versions of each covariate in the regression model.

\begin{figure}
\includegraphics[width=6.5in]{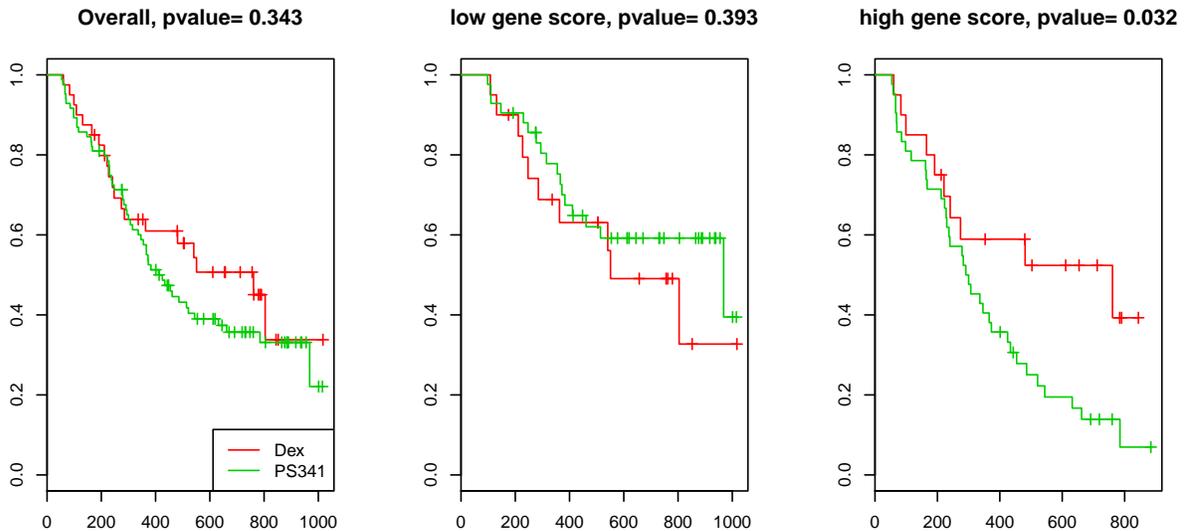}
\caption{\em Example of the modified covariate approach,
applied to gene expression data from multiple myeloma patients
who were given one of two treatments in a randomized trial.
Our procedure constructed a gene score based on 20 genes, to detect
gene expression- treatment interactions. The numerical score was constructed on a training set,
and then categorized into low, medium and high. The  panels show the
 survival curves for a separate test set, overall and  stratified by
the score.}
\label{fig:MMsurv}
\end{figure}

Figure \ref{fig:MMsurv} gives a preview of the results of our method.
The data consist of gene expression measurements from multiple myeloma patients,
who were randomized to one of two treatments.
Our proposed method  constructs a numerical gene score  on a training set  to
reveal  gene expression- treatment interactions. The panels show the
 estimated survival curves for  patients in a
separate test set, overall and  stratified by
the score. Although there is no significant survival difference between
the treatments overall, we see that patients with medium and high gene scores
have better survival with treatment PS341 than those with Doxyrubicin.

 In section \ref{sec:proposed},
we describe the methods for continuous, binary as well as survival type
of outcomes. We also establish a simple casual interpretation of
the proposed method in several cases. In section 3, the finite sample
performance of the proposed method has been investigated via extensive
numerical study. In section 4, we apply the proposed method to a real
data example about the Tamoxifen treatment for breast cancer patients. Finally, potential extensions and applications
of the method were discussed in section 5.

\section{The proposed method}
\label{sec:proposed}
 In the following, we let $T=\pm 1$ be the binary treatment indicator and $Y^{(1)}$ and $Y^{(-1)}$ be the potential outcome if the patient received treatment  $T=1$ and $-1$, respectively. We only observe $Y=Y^{(T)},$ $T$ and $\bZ$, a $q-$dimensional baseline covariate vector. We assume that the observed data consist
of $N$ independent and identically distributed copies of $(Y, T, \bZ),$
$\{(Y_i, T_i, \bZ_i), i=1, \cdots, N\}.$  Furthermore, we let $\bW(\cdot): R^q \rightarrow R^p$ be a $p$ dimensional functions of baseline covariates $\bZ$ and always include an intercept. We denote $\bW(\bZ_i)$ by $\bW_i$ in the rest of the paper. Here the dimension of $\bW_i$ could be large relative to the sample size $N.$
For simplicity, we assume that
$\mbox{Prob}(T=1)=\mbox{Prob}(T=-1)=1/2.$
\subsection{Continuous response model}
When $Y$ is continuous response, a simple multivariate linear regression
model for characterizing the interaction between treatment and covariates is
\begin{equation} Y=\beta_0'\bW(\bZ)+\bgamma_0'\bW(\bZ) \cdot T/2+\epsilon,
\label{multlinear} \end{equation} where $\epsilon$ is the mean zero
random error. In this simple model, the interaction
term $\bgamma_0'\bW(Z) \cdot T$ models the heterogeneous treatment effect
across the population and the linear combination of $\bgamma_0'\bW(\bZ)$
can be used for identifying the subgroup of patients who may or may not
be benefited from the treatment. Specifically, under model (\ref{multlinear}),
we have
\begin{eqnarray*}
\Delta(\bz)&=&{\rm E}(Y^{(1)}-Y^{(-1)}|\bZ=\bz)\\
           &=&{\rm E}(Y|T=1, \bZ=\bz)-{\rm E}(Y|T=-1, \bZ=\bz)\\
           &=& \bgamma_0'\bW(\bz),
\end{eqnarray*} i.e.,
$\bgamma_0'\bW(\bz)$ measures the causal treatment effect for patients with
baseline covariate $\bZ.$  With observed data, $\bgamma_0$ can be
estimated along with $\beta_0$ via the ordinary least squares method.

  On the other hand, noting
the relationship that $$ {\rm E}(2YT|\bZ=\bz)=\Delta(\bz),$$ one may
estimate $\bgamma_0$ by directly minimizing \begin{equation} N^{-1}\sum_{i=1}^N
(2Y_iT_i-\bgamma'\bW_i)^2. \label{causalobj} \end{equation}
We call this the {\em modified outcome} method, where $2YT$ can be viewed as the {\em modified outcome},which has been first proposed in Ph.D thesis of James Sinovitch, Harvard University.

Under the simple linear model (\ref{multlinear}), both estimators are consistent for $\bgamma_0,$ and the full least squares
approach in general is more efficient than the modified outcome method.
In practice, the simple multivariate linear regression model often
is just a working model approximating the complicated underlying
probabilistic relationship between the treatment, baseline
covariates and outcome variables.  It comes as a surprise, that even when model (\ref{multlinear})
is misspecified,  multivariate linear regression and modified outcome estimators still converge to the same deterministic limit $\bgamma^*$ and furthermore $\bW(\bz)'\bgamma^*$ is still a sensible estimator for the interaction effect in the sense that it seeks the ``best'' function of $\bz$ in a functional space ${\cal F}$ to approximate $\Delta(\bz)$ by solving the optimization problem:
 $$ \min_f {\rm E} \{\Delta(\bZ)-f(\bZ)\}^2,$$
 $$\mbox{subject to } f\in {\cal F}=\{\bgamma'\bW(\bz)| \bgamma\in R^p\},$$
where the expectation is with respect to $\bZ.$

\subsection{The Modified Covariate Method}
The modified outcomes estimator defined above is useful for the Gaussian case,
but does not generalize easily to more complicated models.
Hence we propose a new estimator which is equivalent to
the modified outcomes approach in the Gaussian case and
extends easily to other models.
This is the main proposal of this paper.

We consider the simple working model \begin{equation}
Y=\alpha_0+\bgamma_0'\frac{\bW(\bZ)\cdot T}{2}+\epsilon,
\label{proposal} \end{equation} where $\epsilon$ is the mean zero
random error. Based on model (\ref{proposal}), we propose the {\em modified covariate}  estimator $\hat{\bgamma}$
as the minimizer of
\begin{eqnarray}
 \frac{1}{N}\sum_{i=1}^N \left(Y_i-\bgamma'\frac{\bW_i\cdot
T_i}{2}\right)^2.
\label{eqn:modcov}
\end{eqnarray}
The fact that we can directly estimate $\bgamma_0$ in model (\ref{proposal}) without considering the intercept $\alpha_0$ is due to the orthogonality between $\bW({\bZ}_i)\cdot T_i$ and the intercept, which is the consequence of the randomization.
That is, we simply multiply each component of $\bW_i$ by one-half the treatment assignment indicator ($=\pm 1)$ and perform a regular linear regression.
Now since
 $$\frac{1}{N}\sum_{i=1}^N \left\{Y_i-\bgamma'\frac{\bW_i\cdot
T_i}{2}\right\}^2=\frac{1}{4N}\sum_{i=1}^N
\left\{2Y_iT_i-\bgamma'\bW_i\right\}^2,$$
the modified outcome and modified covariate estimates
are identical and share the same causal interpretation for the simple Gaussian model.  Operationally, we can  omit the intercept and perform a simple linear regression with the modified covariates.  In general, we proposed the following modified covariate approach
$$~$$
\fbox{
\begin{minipage}{\textwidth}
\begin{enumerate}
\item Modify the covariate
$$Z_i \rightarrow \bW_i=\bW(\bZ_i) \rightarrow \bW_i^*=\bW_i\cdot T_i/2$$
\item Perform appropriate regression
 \begin{equation}
Y \sim \bgamma_0'\bW^*
\label{eqn:proposal}
\end{equation}
based on the modified observations
\begin{eqnarray}
 (\bW^*_i, Y_i)=\{(\bW_i\cdot T_i)/2, Y_i\}, i=1,2,\ldots N.
\label{eqn:moddata}
\end{eqnarray}
\item $\hat{\bgamma}'\bW(\bz)$ can be used to stratify patients for individualized treatment selection.
\end{enumerate}
\end{minipage}
}
$$~$$

\begin{figure}
\begin{center}
\includegraphics[width=5.5in]{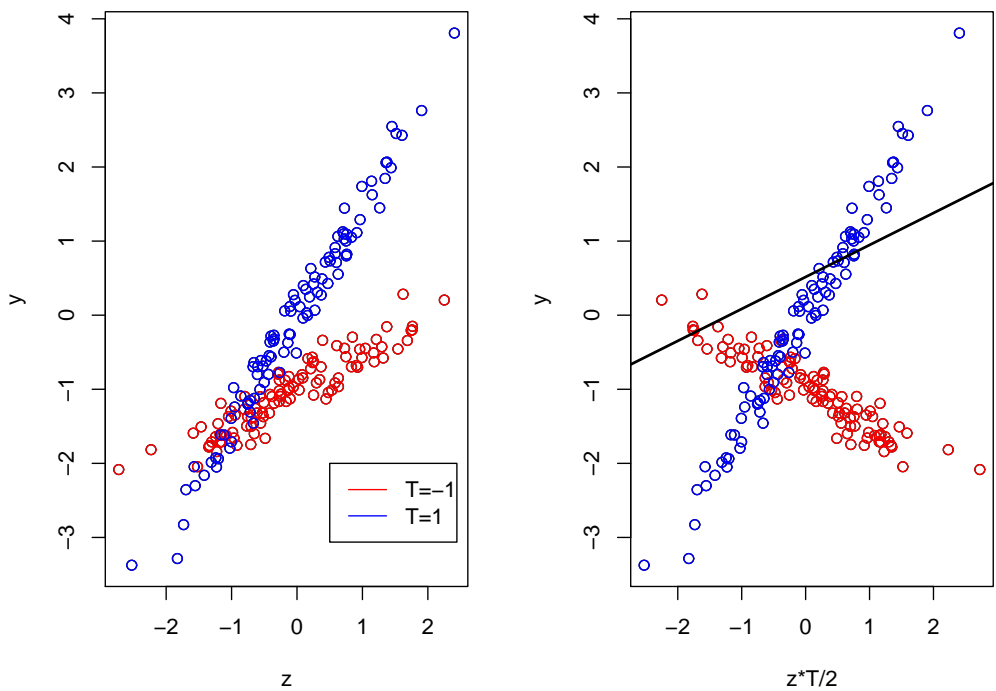}
\end{center}
\caption{\em Example of the modified covariate approach.
The raw data is shown the left, consisting of a single covariate
$Z$ and a treatment $T=-1$ or $1$.
The treatment-covariate interaction has slope $\gamma$ approximately
equal to 1.
On the right panel we have plotted the response against $Z\cdot T/2$.
The the regression line computed in the right panel estimates the treatment effect for each give value of covariate $Z.$}
\label{fig:example}
\end{figure}

Figure \ref{fig:example} illustrates how  the  modified covariate method works
for a single covariate $Z$,
in two treatment groups.
The raw data is shown the left, and the data with  modified covariate
is shown on the right.
The slope of the regression line computed in the right panel   estimates the
treatment-covariate interaction.

The advantage of this new approach
is twofold: it avoids having to directly model the main effects and it has a
causal interpretation for the resulting estimator regardless of the adequacy of the assumed working model (\ref{proposal}).
Furthermore, unlike modified outcome method, it is straightforward to generalize the new approach to other types of outcome.

\subsection{Binary Responses} When $Y$ is a binary response, in the
same spirit as the  continuous outcome case, we propose to fit
a multivariate logistic regression model with modified covariates $\bW^*=\bW(\bZ)\cdot T/2$ generalized from (\ref{eqn:proposal}):
\begin{equation} \mbox{Prob}(Y=1|\bZ,
T)=\frac{\exp(\bgamma_0' \bW^*)}{1+\exp(\bgamma_0'
\bW^*)}.  \label{propbinary} \end{equation}
Noting that if model (\ref{propbinary}) is correctly specified,
then
\begin{eqnarray*}
\Delta(\bz)&=&{\rm Prob}(Y^{(1)}=1|\bZ=\bz)-{\rm Prob}(Y^{(-1)}=1|\bZ=\bz)\\
           &=&{\rm Prob}(Y=1|T=1, \bZ=\bz)-{\rm Prob}(Y=1 |T=-1, \bZ=\bz)\\
           &=& \frac{\exp\{\bgamma_0'\bW(\bz)/2\}-1}{\exp\{\bgamma_0' \bW(\bz)/2\}+1},
\end{eqnarray*}
and thus $\bgamma_0'\bW(\bz)$ has an appropriate causal interpretation.
However, even when model (\ref{propbinary}) is not correctly specified,
we still can estimate $\bgamma_0$ by treating (\ref{propbinary}) as a working model.

In general, the maximum likelihood estimator (MLE) of the working model, converges
to a deterministic limit $\bgamma^*$ and $\bW(\bz)'\bgamma^*/2$ can be viewed as the solution to the following optimization problem
$$ \mbox{max}_f \mbox{E} \left\{ Yf(\bZ)T-\log(1+e^{f(\bZ)T})\right\} $$
$$ \mbox{subject to } f\in {\cal F}=\{\gamma'\bW(\bz)/2| \bgamma\in R^p \},$$
where the expectation is with respect to $(Y, T, \bZ).$
Therefore, where $\bW(\bz)$ forms a ``rich" set of basis functions, $\bW(\bz)'\bgamma^*/2$ is an approximation to the minimizer of $\mbox{E} \left\{ Yf(\bZ)T-\log(1+e^{f(\bZ)T})\right\}.$ In the appendix, we show that the latter can be represented as
$$f^*(\bz)=\log\left\{\frac{1-\Delta(\bz)}{1+\Delta(\bz)}  \right\}$$ under very general assumptions. Therefore,
$$\hat{\Delta}(\bz)=\frac{\exp\{\hat{\bgamma}'\bW(\bz)/2\}-1}{\exp\{\hat{\bgamma}'\bW(\bz)/2\}+1}$$ may serve as an estimate for the covariate-specific treatment effect and used to stratify patients population, regardless of the validity the working model assumptions.

 As described above, the MLE from the working model (\ref{propbinary}) can always be used to construct a surrogate to the personalized treatment effect measured by the ``risk difference''
 $$\Delta(\bz)=\mbox{E}(Y^{(1)}-Y^{(-1)}|\bZ=\bz).$$
 On the other hand,  different measures for individualized treatment effects such as relative risk may also be of interest. For example, if we consider an alternative approach for fitting the logistic regression working model (\ref{propbinary}) by letting
$$\hat{\bgamma}=\mbox{argmax}_{\bgamma} \sum_{i=1}^n \left\{(1-Y_i)\bgamma'\bW^*-Y_ie^{-\bgamma'\bW^*_i}\right\},$$
 then $\hat{\bgamma}$ converges to a deterministic limit $\tilde{\bgamma}^{*}$ and $\bW(\bz)'\tilde{\bgamma}^{*}(\bz)/2$ can be viewed as an approximation to $\log\{\tilde{\Delta}(\bz)\},$ where
$$ \tilde{\Delta}(\bz)=\frac{\mbox{Prob}(Y^{(1)}=1|\bZ=\bz)}{\mbox{Prob}(Y^{(-1)}=1|\bZ=\bz)},$$
which measures the treatment effect based on ``relative risk" rather than ``risk difference''. The detailed justification is given in the Appendix 6.1.

\subsection{Survival Responses} When the outcome variable is survival
time, we often do not observe the exact outcome for every subject
in a clinical study due to incomplete follow-up. In this case,
we assume that the outcome $Y$ is a pair of random variables $(X,
\delta)=\{\tilde{X}\wedge C, I(\tilde{X}<C)\},$ where $\tilde{X}$ is the
survival time of primary interest, $C$ is the censoring time and $\delta$
is the censoring indicator.

Firstly, we propose to fit a Cox regression model
\begin{equation} \lambda(t|\bZ, T)=\lambda_0(t)e^{\bgamma'\bW^*} \label{propsurv2} \end{equation}
where $\lambda(t|\cdot)$ is
the hazard function for survival time $\tilde{X}$ and $\lambda_0(\cdot)$
is a baseline hazard function free of $\bZ$ and $T.$
When model (\ref{propsurv2}) is correctly specified,

\begin{eqnarray*}
\Delta(\bz)&=&\log\left[\frac{\mbox{E} \{\Lambda_0(\tilde{X}^{(1)})| \bZ=\bz\}}{\mbox{E}\{\Lambda_0(\tilde{X}^{(-1)})|\bZ=\bz\}}\right]\\
 &=&\left[\frac{\mbox{E} \{\Lambda_0(\tilde{X})| T=1, \bZ=\bz\}}{\mbox{E}\{\Lambda_0(\tilde{X})| T=-1, \bZ=\bz\}}\right]\\
&=& \exp\{-\bgamma_0'\bW(\bz)\}
\end{eqnarray*}
and $\bgamma_0'\bW(\bz)$ can be used to stratify patient population according to $\Delta(\bz),$
where $\Lambda_0(t)=\int_0^t \lambda_0(u)du$ is a monotone increasing function (the baseline cumulative hazard function).
Under the proportional hazards assumption, the maximum partial likelihood estimator $\hat{\bgamma}$ is a consistent estimator for $\bgamma_0$ and semiparametric efficient.
Moreover, even when model (\ref{propsurv2}) is misspecified,
we still can ``estimate'' $\bgamma_0$ by maximizing the partial likelihood
function. In general, the resulting estimator, $\hat{\bgamma},$ converges to a deterministic limit
$\bgamma^*$, which is the root of a limiting score equation \citep{Lin:Wei:1989}.
More generally, $\bW(\bz)'\bgamma^*/2$ can be viewed as the solution of the optimization problem
$$  \max_f \mbox{E} \int_0^\tau \left[f(\bZ)T- \log\{\sum_{j=1}^N e^{f(\bZ)T}I(\tilde{X}\ge u) \} \right]d N(u)$$
$$  \mbox{subject to } f\in {\cal F}=\{\bgamma'\bW(\bz)/2| \bgamma \in R^p\},$$
where $N(t)=I(\tilde{X}\le t)\delta_i$ and the expectation is with respect to $(Y, T, \bZ).$ Therefore, $\bW(\bz)'\bgamma^*/2$ can be viewed as an approximation to
$$f^*(\bz)=\mbox{argmax}_f \mbox{E} \int_0^\tau \left[f(\bZ)T- \log\{\sum_{j=1}^N e^{f(\bZ)T}I(\tilde{X}\ge u) \} \right]d N(u).$$
In appendix 6.1, we shown that the minimizer $f^*$ satisfies
$$  e^{f^*(\bz)}\mbox{E}\{\Lambda^*(\tilde{X}^{(1)})|\bZ=\bz\}-e^{-f^*(\bz)}\mbox{E}\{\Lambda^*(\tilde{X}^{(-1)})|\bZ=\bz\}=\mbox{E}(\Delta^{(1)}|\bZ=\bz)-\mbox{E}(\Delta^{(-1)}|\bZ=\bz)$$
for a monotone increasing function $\Lambda^*(u).$ Thus, when censoring rates are balanced between two arms,
$$f^*(\bz)\approx -\frac{1}{2}\log \left[ \frac{\mbox{E} \{\Lambda^*(\tilde{X}^{(1)})|\bZ=\bz\}}{\mbox{E}\{\Lambda^*(\tilde{X}^{(-1)})|\bZ=\bz\}}\right]$$
can be used for characterizing the covariate-specific treatment effect and stratifying the patient population even when the working model (\ref{propsurv2}) is misspecified.

\subsection{Regularization for high dimensional data}

When the dimension of $\bW^*$, $p,$ is high, we can
easily apply appropriate variable selection procedures based
on the corresponding working model. For example, $L_1$ penalized
(Lasso) estimators proposed by \citet{Ti96} can be
directly applied to the modified data (\ref{eqn:moddata}). In general,
one may estimate $\bgamma$ by minimizing
\begin{equation} \frac{1}{N}\sum_{i=1}^N l(Y_i, \bgamma'\bW^*_i)+\lambda_0\sum_{j=1}^p |\gamma_{j}|,
\label{eqn:lasso}
\end{equation}
where
$$
l(Y_i, \bgamma'\bW^*_i)=\begin{cases} \frac{1}{2}(Y_i-\bgamma'\bW^*_i)^2 &\mbox{ for continuous response}\\
                                      -\{Y_i\bgamma'\bW^*_i-\log(1+e^{\bgamma'\bW^*_i})\} &\mbox{ for binary response}\\
                                      -\left[\bgamma'\bW^*_i- \log\{\sum_{j=1}^N e^{\bgamma'\bW^*_i}I(X_j\ge X_i) \} \right]\Delta_i &\mbox{ for survival response}
                                              \end{cases}. $$

It might be  reasonable to  suppose that the covariates interacting with the treatment will more likely
be the ones exhibiting important main effects themselves. Therefore,
one could also apply the adaptive Lasso procedure \citep{Zou2006a}
with feature weights $\hat{w}_j$ proportional to  the reciprocal  of the univariate ``association strength''
between the outcome $Y$ and the $j$th component of $\bW(\bZ).$
 Specifically, one may modify the penalty in (\ref{eqn:lasso}) as
\begin{equation} \lambda_0\sum_{j=1}^p\frac{|\gamma_{j}|}{\hat{w}_j},
\label{eqn:adaplasso}
\end{equation}
where $\hat{w}_j=|\hat{\theta}_i|^{-1}$ or $(|\hat{\theta}_{-1i}|+|\hat{\theta}_{1i}|)^{-1},$ where $\hat{\theta}_{j1},$ $\hat{\theta}_{j(-1)}$ and $\hat{\theta}_{j},$  are the estimated regression coefficients of the $j$th component of $\bW(\bZ)$ in appropriate univariate regression analysis with observations from the group $T=1$ only, from the group $T=-1$ only, and from both groups, respectively. Other regularization methods such as elastic net may also be used \citep{Zou:Hastie:2005}.

Interestingly, one can treat the modified data (\ref{eqn:moddata})
just as generic data and hence couple it with other statistical learning techniques. For example, one can apply a classifier such as  prediction analysis of microarrays (PAM) to the modified data for the purpose of finding subgroup of samples in which the treatment effect is large.  We also can do large scale hypothesis testing
on the  modified data to determine which gene-treatment interactions  have a significant
effect on the outcome.

\subsection{Efficiency Augmentation}
When the models (\ref{eqn:proposal}, \ref{propbinary} and \ref{propsurv2}) with modified covariates is correctly specified, the MLE estimator for $\bgamma^*$ is the most efficient estimator asymptotically. However, when models are treated as working models subject to mis-specification, a more efficient estimator can be obtained for estimating the same deterministic limit $\bgamma^*.$ To this end, noting the fact that in general $\hat{\bgamma}$ is defined as the minimizer of an objective function motivated from a working model:
\begin{equation}
\hat{\bgamma}=\mbox{argmin}_{\bgamma} \frac{1}{N}\sum_{i=1}^N l(Y_i, \bgamma'\bW^*_i)
\end{equation}\label{obj}
Noting that for any function $\ba(\bz): R^q \rightarrow R^p,$  $E\{T_i \ba(\bZ_i)\}=0$ due to randomization, the minimizer of the augmented objective function
$$\frac{1}{N}\sum_{i=1}^N \left\{ l(Y_i, \bgamma'\bW^*_i)-T_i \ba(\bZ_i)'\bgamma \right\} $$
converges to the same limit as $\hat{\bgamma},$ when $N \rightarrow \infty.$ Furthermore, by selecting an optimal augmentation term $\ba_0(\cdot)$, the minimizer of the augmented objective function can have smaller variance than that of the minimizer of the original objective function.

In appendix 6.2, we show that $$\ba_0(\bz)=-\frac{1}{2}\bW(\bz) \mbox{E}(Y|\bZ=\bz)$$
 and $$\ba_0(\bz)=-\frac{1}{2}\bW(\bz) \{\mbox{E}(Y|\bZ=\bz)-0.5\}$$ are optimal choices for continuous and binary responses, respectively. Therefore, we proposed the following two-step procedures for estimating $\bgamma^*:$
$$~$$
\fbox{
\begin{minipage}{\textwidth}
\begin{enumerate}
\item Estimate the optimal $\ba_0(\bz):$
      \begin{enumerate}
      \item For continuous response, fit the linear regression model $E(Y|\bZ)=\xi'B(\bZ)$ for appropriate function $B(\bZ)$ with OLS. Appropriate regularization will be used if the dimension of $B(\bZ)$ is high.  Let
       $$\hat{\ba}(\bz)=-\frac{1}{2}\bW(\bz)\times \hat{\xi}'B(\bz).$$
       \item For binary response, fit the logistic regression model $\mbox{logit}\{\mbox{Prob}(Y=1|\bZ)\}=\xi'B(\bZ)$ for appropriate function $B(\bZ)$ by maximizing the likelihood function. Appropriate regularization will be used if the dimension of $B(\bZ)$ is high.  Let
       $$\hat{\ba}(\bz)=-\frac{1}{2}\bW(\bz) \times \left\{\frac{e^{\hat{\xi}'B(\bz)}}{1+e^{\hat{\xi}'B(\bz)}}-\frac{1}{2}\right\}.$$
      \end{enumerate}
      Here $B(\bz)=\{B_1(\bz), \cdots, B_S(\bz)\}$ and $B_k(\bz): R^q \rightarrow R^1 $ is selected basis function.
\item Estimate $\bgamma^*$
    \begin{enumerate}
      \item For continuous response, we minimize $$\frac{1}{N}\sum_{i=1}^N \left\{\frac{1}{2}(Y_i-\bgamma'\bW^*_i)^2-\bgamma'\hat{\ba}(\bZ_i)T_i\right\}$$
           with appropriate regularization if needed.
       \item For binary response, we minimize $$\frac{1}{N}\sum_{i=1}^N \left[-\{Y_i\bgamma'\bW^*_i-\log(1+e^{\bgamma'\bW^*_i})\}-\bgamma'\hat{\ba}(\bZ_i)T_i\right]$$
           with appropriate regularization if needed.
      \end{enumerate}
\end{enumerate}
\end{minipage}
}
$$~$$

For survival outcome, the log-partial likelihood function is not a simple sum of i.i.d terms. However, in Appendix 6.2 we show that the optimal choice of $\ba(\bz)$ is
$$\ba_0(\bz)=-\frac{1}{2}\left[\frac{1}{2}\bW(\bz)\left\{G_1(\tau; \bz)+G_2(\tau; \bz)\right\}-\int_0^\tau \mathbf{R}(u) \{G_1(du; \bz)-G_2(du; \bz)\}\right],$$
where $G_1(u; \bz)=E\{M(u)|\bZ=\bz, T=1\},$
$G_2(u;\bz)=E\{M(u)|\bZ=\bz, T=-1\},$
$$M(t, \bW^*, \bgamma^*)=N (t)-\int_0^t \frac{I(X \ge u) e^{\bgamma'\bW^*} d {\rm E}\{N(u)\}}{{\rm E}\{e^{\bgamma'\bW^*}I(X\ge u)\}}$$
and
$$\mathbf{R}(u; \bgamma^*)=\frac{{\rm E}\{\bW^*e^{\bgamma'\bW^*}I(X\ge u)\}}{{\rm E}\{e^{\bgamma'\bW^*}I(X\ge u)\}}.$$
Unfortunately, $\ba_0(\bz)$
depends on the unknown parameter $\bgamma^*.$ On the other hand, on
high-dimensional case, the interaction effect is usually small and
it is not unreasonable to assume that $\bgamma^*\approx 0.$ Furthermore,
if the censoring patterns are similar in both arms, we have $G_1(u,
\bz)\approx G_2(u, \bz).$ Using these two approximations, we can
simplify the optimal augmentation term as
$$\ba_0(\bz)=-\frac{1}{4}\bW(\bz)\left\{G_1(\tau; \bz)+G_2(\tau; \bz)\right\}=-\frac{1}{2}\bW(\bz) \times \mbox{E}\{M(\tau)|\bZ=\bz)$$
where
$$M(t)=N (t)-\int_0^t \frac{I(X \ge u) d {\rm E}\{N(u)\}}{{\rm E}\{I(X\ge
u)\}}.$$
 Therefore,  we propose to employ the following approach for implementing the efficiency augmentation procedure,:\\

\fbox{
\begin{minipage}{\textwidth}
\begin{enumerate}
\item Calculate
        $$\hat{M}_i(\tau)=N_i(\tau)-\int_0^{\tau}\frac{I(X_i\ge u) d \{\sum_{j=1}^N N_j(u)\}}{\sum_{j=1}^N I(X_j\ge u)}$$
      for $i=1, \cdots, N$ and fit the linear regression model $E(\hat{M}(t)|\bZ)=\xi'B(\bZ)$ for appropriate function $B(\bZ)$ with OLS and appropriate regularization if needed.  Let
       $$\hat{\ba}(\bz)=-\frac{1}{2}\bW(\bz)\times \hat{\xi}'B(\bz).$$
\item Estimate $\bgamma^*$ by minimizing
    $$\frac{1}{N}\sum_{i=1}^N \left(-\left[\bgamma'\bW^*_i- \log \{\sum_{j=1}^N e^{\bgamma'\bW^*_i}I(X_j\ge X_i)\} \right]\Delta_i-\bgamma'\hat{\ba}(\bZ_i)T_i\right)$$
    with appropriate penalization if needed.
\end{enumerate}
\end{minipage}
}
$$~$$
\noindent
{\bf Remarks 1}

When the response is continuous, the efficient augmentation estimator is the minimizer of
\begin{align*}
& \sum_{i=1}^N
\left[\frac{1}{2}\left\{Y_i-\frac{1}{2}\bgamma'\bW(\bZ_i)T_i/2\right\}^2-\bgamma'\hat{\ba}(\bZ_i)T_i \right] \\
=&
\sum_{i=1}^N
\frac{1}{2}\left\{Y_i-\hat{\xi}'B(\bZ_i)-\frac{1}{2}\bgamma'\bW(\bZ_i)T_i \right\}^2+\mbox{constant}.
\end{align*}
This equivalence implies that this efficiency augmentation
procedures is asymptotically equivalent to that based on a simple
multivariate regression with main effect $\hat{\xi}'B(\bZ_i)$ and
interaction $\bgamma'\bW(\bZ)\cdot T.$  This is not a surprise. As we pointed out in section 2.1, the choice of the main effect in the linear regression does not affect the asymptotical consistency of estimating the interactions. On the other hand, a good choice of main effect model can help to estimate the interaction, i.e., personalized treatment effect, more accurately.

Another consequence is that one may directly use the same algorithm
solving standard optimization problem to obtain the augmented estimator when lasso
penalty is used. For binary or survival response, the
augmented estimator under lasso regularization can be obtained with slightly modified algorithm designed for lasso optimization as well. The detailed algorithm is given in the appendix 6.3.
$$ $$

\noindent
{\bf Remarks 2}

 For nonlinear models such as logistic and Cox regressions, the augmentation method is NOT equivalent to the full regression approach including main effect and interaction terms. In those cases, different specification of the main effects in the regression model result in asymptotically different estimates for the interaction term, which, unlike the proposed modified covariate estimator, in general can not be interpreted as the personalized treatment effect.

$$ $$
\noindent
{\bf Remarks 3}

With binary response, the estimating equation targeting on approximating the relative risk is
$$\sum_{i=1}^N \bW^*_i\{(1-Y_i)-Y_ie^{-\bgamma'\bW^*_i}\}$$
and the optimal augmentation term $a_0(\bz)$ can be be approximated by
$$-\frac{1}{2} \bW(\bz)\left \{\mbox{E}(Y|\bZ=\bz)- \frac{1}{2}\right\}$$
when $\bgamma^* \approx 0.$ The efficiency augmentation algorithm can be carried out accordingly.

$$ $$
\noindent
{\bf Remarks 4}

The similar technique can also be used for improving other estimators such as that proposed by \cite{ZZRK:12}, where the surrogate objective function for the weighted mis-classification error can be written in the form of (\ref{obj}) as well. The optimal function $\ba_0(\bz)$ needs to be derived case by case.

\section{Numerical Studies}
\label{sec:numerical}
 In this section, we perform an extensive
numerical study to investigate the finite sample performance of proposed method in various settings: the treatment may or may not have marginal main effect between two groups; the personalized treatment effect may depend on complicated function of covariates such as interactions among covariates; the regression model for detecting the interaction may or may not be correctly specified.  Due to the limitation of the space, we only present simulation results from the selected representive cases. The results for other scenarios are similar to those presented.

\subsection{Continuous responses}
\begin{figure}[t] \centering
\includegraphics[width=5in]{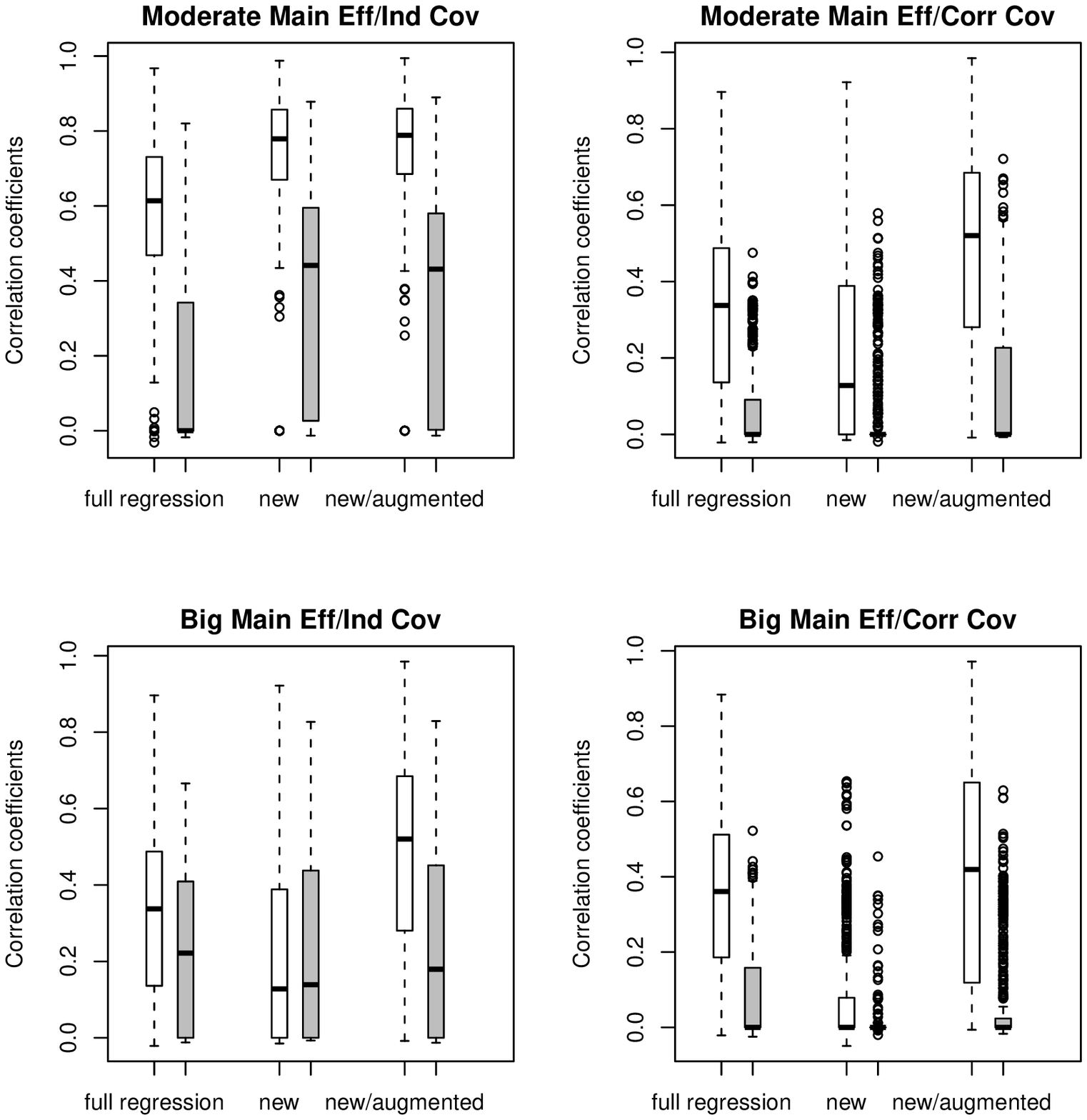}
\caption{\em Boxplots for the correlation coefficients between the estimated score and true treatment effect
with three different methods applied to continuous outcomes. The empty and filled boxes represent low and high dimensional ($p=50$ and $p=1000$) cases, respectively.
Left upper panel: moderate main effect and independent covariates; right upper panel: moderate main effect and correlated covariates; left lower panel: big main effect and independent covariates; right lower panel: big main effect and correlated covariates.}
 \label{fig:gauss}
\end{figure}

For continuous responses, we generated $N$ independent Gaussian samples from the regression model
\begin{eqnarray}
Y&=&\sum_{j=1}^p \beta_j Z_j + \sum_{j=1}^p \gamma_j Z_j T+\sigma_0\cdot \epsilon,
\label{eqn:simmodel}
\end{eqnarray}
where the covariate $(Z_1, \cdots, Z_p)$ follows a mean zero multivariate normal distribution with a compound symmetric variance-covariance matrix, $(1-\rho)\mathbf{I}_p+\rho \mathbf{1}'\mathbf{1},$ and $\epsilon\sim N(0, 1).$  We let $(\gamma_1, \gamma_2, \gamma_3, \gamma_4, \gamma_5, \cdots, \gamma_p)=(1/2, -1/2, 1/2, -1/2, 0, \cdots, 0),$ $\sigma_0=\sqrt{2},$ $N=100,$ and $p=50$ and $1000$ representing high and low dimensional cases, respectively. The treatment $T$ was generated as $\pm 1$ with equal probability at random. We consider four sets of simulations:
\begin{enumerate}
\item  $\beta_j=(-1)^{j+1}I(3\le j\le 10)/4$ and $\rho=0;$
\item  $\beta_j=(-1)^{j+1}I(3\le j\le 10)/4$ and $\rho=0.5;$
\item  $\beta_j=(-1)^{j+1} I(3\le j\le 10)/2$ and $\rho=0;$
\item  $\beta_j=(-1)^{j+1}I(3\le j\le 10)/2$ and $\rho=0.5.$
\end{enumerate}
Settings 1 and 2 presents relative moderate main effect, where the variability in response contributable to the main effect is the same as that to the interaction. Settings 3 and 4 represent relative big main effect, where the variability in response contributable to the main effect is twice as big as that to the interaction.
For each of the simulated data set, we implemented three methods:
\begin{itemize}
\item {\em full regression:} The first method is to fit a multivariate linear regression with full main effect
  and covariate/treatment interaction terms, i.e., the dimension of the
  covariate matrix was $2(p+1)$. The Lasso was used to select the variables.
\item {\em new:} The second method is to fit a multivariate linear regression with the modified covariate $\bW^*=(1, \bZ)'\cdot T/2$ as the covariates, i.e., the dimension of the covariate matrix is $p+1.$ Again, the Lasso is used for selecting variables having treatment interaction.
\item {\em new/augmented:} the proposed method with efficiency augmentation, where $\mbox{E}(Y|\bZ)$ is estimated with lasso-regularized ordinary least squared method and $B(\bz)=\bz.$
\end{itemize}
 For all three methods, we selected the Lasso penalty parameter via 20-fold cross-validation. To evaluate the performance of the resulting score measuring the  individualized treatment effect, we estimated the Spearman's rank correlation coefficient between the estimated score and the ``true'' treatment effect $$\Delta(\bZ)=\mbox{E}(Y^{(1)}-Y^{(-1)}|\bZ)=(Z_1-Z_2+Z_3-Z_4)$$
 in an independently generated set with a sample size of 10000.
Based on 500 sets of simulations, we plotted the boxplots of the rank correlation coefficients between the estimated scores $\hat{\bgamma}'\bZ$ and $\Delta(\bZ)$ under simulation settings (1), (2), (3) and (4) in top left, top right, bottom left and bottom right panels of Figure \ref{fig:gauss}, respectively. When the main effect is moderate and covariates are independent (setting 1),  the performance of the proposed method is better than that of the full regression approach. However, when the main effect is relatively big compared to interactions (settings 3 and 4), the proposed method is unable to estimate the correct individualized treatment effect well and is actually inferior to the simple regression method. On the other hand, the performance of the ``new/augmented'' is the best or nears the best across all the four settings and is sometimes substantially better than its competitors.

\subsection{Binary responses}
For binary responses, we used the same simulation design as that for the continuous response. Specifically, we generated $N$ independent binary samples from the regression model
\begin{eqnarray}
Y&=& I\left(\sum_{j=1}^p \beta_j Z_j + \sum_{j=1}^p \gamma_j Z_j T+\sigma_0\cdot \epsilon \ge 0\right),
\label{eqn:binmodel1}
\end{eqnarray}
where all the model parameters were the same as those in the case of continuous response. Noting that the logistic regression model is misspecified under the chosen simulation design. We also considered the same four settings with different combinations of $\beta_j$ and $\rho.$ For each of the simulated data set, we implemented three methods:
\begin{enumerate}
\item {\em full regression:} The first method is to fit a multivariate logistic regression with full main effect
  and covariate/treatment interaction terms, i.e., the dimension of the
  covariate matrix was $2(p+1)$. The Lasso was used to select the variables.
\item {\em new:} The second method is to fit a multivariate logistic regression (without intercept) with the modified covariate $\bW^*=(1, \bZ)'\cdot T/2$ as the covariates. Again, the Lasso was used for selecting variables having
treatment interaction.
\item {\em new/augmented:} the proposed method with efficiency augmentation, where $\mbox{E}(Y|\bZ)$ is estimated with Lasso-penalized logistic regression.
\end{enumerate}
 To evaluate the performance of the resulting score measuring the  individualized treatment effect, we estimated the Spearman's rank correlation coefficient between the estimated score and the ``true'' treatment effect
 \begin{align*}
 \Delta(\bZ)&=\mbox{E}(Y^{(1)}-Y^{(-1)}|\bZ)\\
            &=\Phi\left(\frac{\sum_{j=1}^p(\beta_j+\gamma_j) Z_j}{\sigma_0}\right)-\Phi\left(\frac{\sum_{j=1}^p(\beta_j-\gamma_j) Z_j}{\sigma_0}\right)
 \end{align*}
 where $\Phi$ was the cumulative distribution function of standard normal. Although the scores measuring the interaction from the first and second/third methods were different even when the sample size goes to infinity, the rank correlation coefficients put them on the same footing in comparing performances.

\begin{figure}[t] \centering
\includegraphics[width=5in]{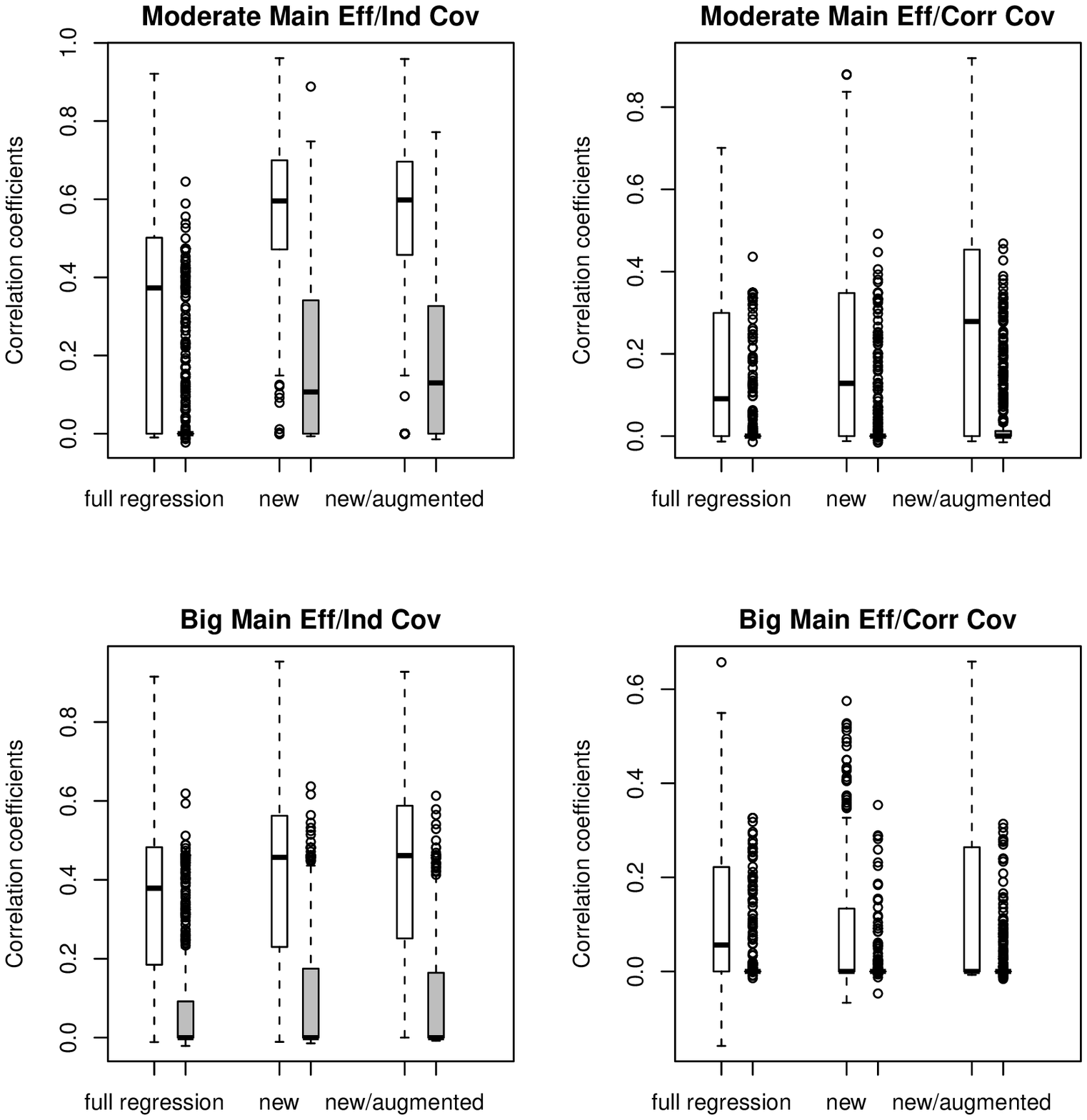}
\caption{\em Boxplots for the correlation coefficients between the estimated score and true treatment effect
with three different methods applied to binary outcomes. The empty and filled boxes represent low and high dimensional ($p=50$ and $p=1000$) cases, respectively.
Left upper panel: moderate main effect and independent covariates; right upper panel: moderate main effect and correlated covariates; left lower panel: big main effect and independent covariates; right lower panel: big main effect and correlated covariates.}
 \label{fig:bin}
\end{figure}

In top left, top right, bottom left and bottom right panels of Figure \ref{fig:bin}, we plotted the boxplots of the correlation coefficients between the estimated scores $\hat{\bgamma}'\bZ$ and $\Delta(\bZ)$ under simulation settings (1), (2), (3) and (4), respectively. The patterns are similar to that for the continuous response. The ``new/augmented method'' performed the best or close to the best in all the four settings. The efficiency gain of the augmented method in setting 4 where the main effect was relative big and covariates were correlated was more significant than that in other settings.

\begin{figure}
\begin{center}
\includegraphics[width=3.5in]{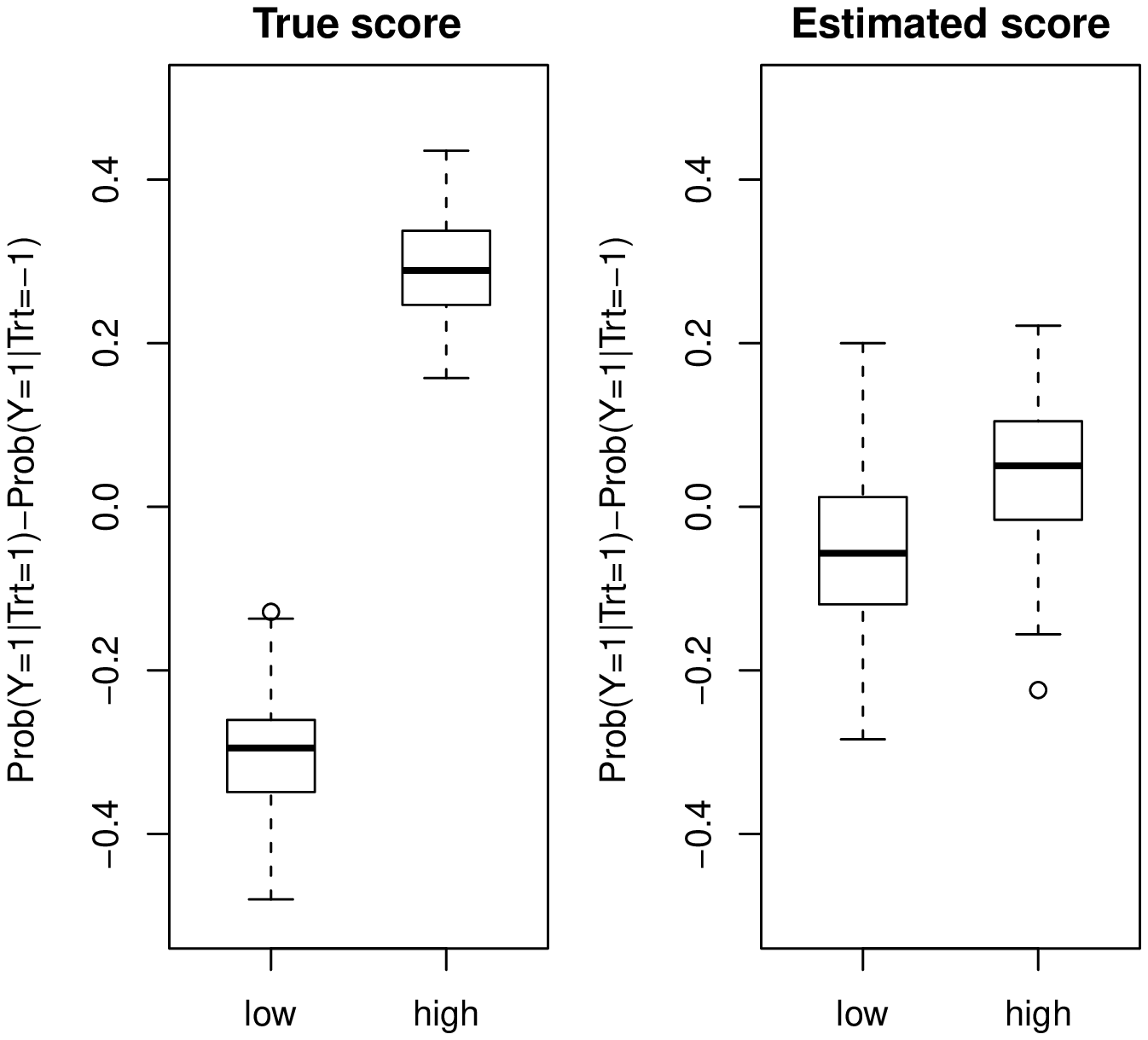}
\end{center}
\caption{Left panel: the boxplots for the group differences in $Y$ in subgroups stratified by the optimal score; right panel: the boxplots for the group differences in $Y$ in subgroups stratified by posterior probability based on the independently trained nearest shrunken centroid classifier.}
\label{fig:logist}
\end{figure}

 In additional simulation study, we also evaluated the empirical performance of the generalized modified covariate approach with nearest shrunken centroid classifier. In one set of the simulation, the binary response is simulated from model (\ref{eqn:binmodel1}) with $p=50$, $n=200$, $\beta_j=I(j\le 20)/2,$ $\gamma_j=I(j\le 4)/2$ and $\sigma_0=\sqrt{2}.$ Here the first four predictors have covariate treatment interaction. We applied the nearest shrunken centroid classifier \citep{THNC2002} to the modified data (\ref{eqn:moddata}) with the shrinkage parameter selected via 10 fold cross-validation. This produced a posterior probability estimator for $\{Y=1\}.$ We then applied this estimated posterior probability interaction score, to a independently generated test set of size 400. We dichotomized the observations in the test set into high and low score groups according to the median value and calculated the differences between two treatment arms in high and low score groups separately. With 100 replications, the boxplots of the differences in high and low score groups were shown in the right panel of Figure \ref{fig:logist}. For comparison purposes, the empirical differences between two arms in high and low score groups determined by the true interaction score $\sum_{j=1}^p \gamma_jZ_j$ were shown in the left panel of figure \ref{fig:logist}. It can be seen that modified covariate approach, coupled with nearest shrunken centroid classifier, provided reasonable stratification for differentiating the treatment effect.

\subsection{Survival Responses}
For survival responses, we used the same simulation design as that for the continuous and binary responses. Specifically, we generated $N$ independent survival time from the regression model
\begin{eqnarray}
\tilde{X}&=& \exp\left(\sum_{j=1}^p \beta_j Z_j + \sum_{j=1}^p \gamma_j Z_j T+\sigma_0\cdot \epsilon\right),
\label{eqn:survmodel1}
\end{eqnarray}
where all the model parameters were the same as in the previous subsections. The censoring time was generated from uniform distribution $U(0, \xi_0),$ where $\xi_0$ was selected to induce 25\% censoring rate. For each of the simulated data set, we implemented three methods:
\begin{enumerate}
\item {\em full regression:} The first method was to fit a multivariate Cox regression with full main effect
  and covariate/treatment interaction terms, i.e., the dimension of the
  covariate matrix was $2p+2$. The Lasso was used to select the variables
\item {\em new:} The second method was to fit a multivariate Cox regression with the modified covariate $\bW^*=(1, \bZ)'\cdot T/2$ as the covariates. Again, the Lasso was used for selecting variables having
treatment interaction.
\item {\em new/augmented:} the proposed method with efficiency augmentation. To model the $\mbox{E}\{M(\tau)|\bZ\}$, we used linear regression with lasso regularization method.
\end{enumerate}
 To evaluate the performance of the resulting score measuring the  individualized treatment effect, we estimated the Spearman's rank correlation coefficient between the estimated score and the ``true'' treatment effect based on survival probability at $t_0=5$
 \begin{align*}
 \Delta(\bZ)&=\mbox{Prob}(\tilde{X}^{(1)}\ge t_0|\bZ)-\mbox{Prob}(\tilde{X}^{(-1)}\ge t_0|\bZ)\\
            &=\Phi\left(\frac{\sum_{j=1}^p(\beta_j+\gamma_j) Z_j-\log(t_0)}{\sigma_0}\right)-\Phi\left(\frac{\sum_{j=1}^p(\beta_j-\gamma_j) Z_j-\log(t_0)}{\sigma_0}\right).
 \end{align*}

\begin{figure}[t] \centering
\includegraphics[width=5in]{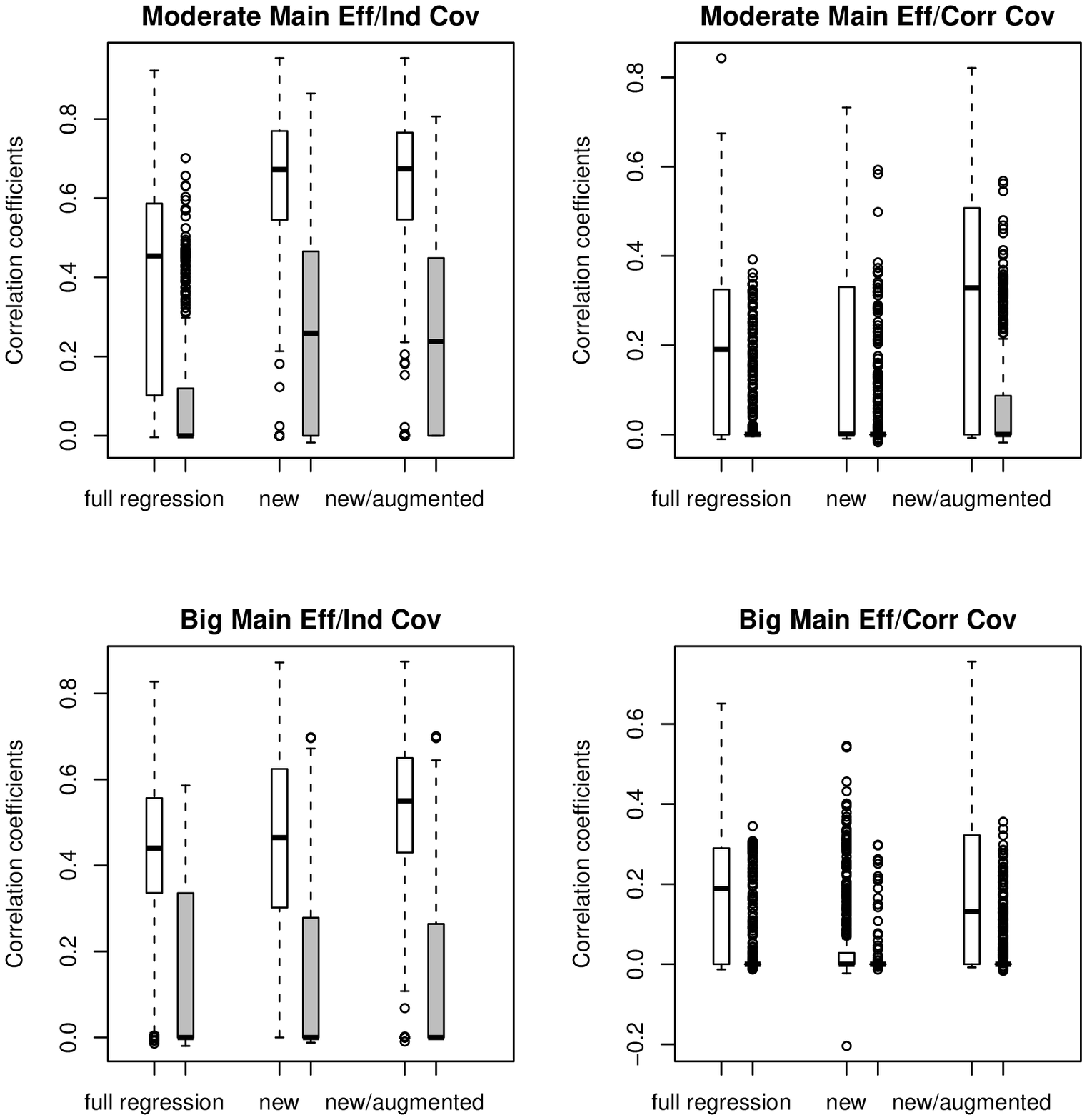}
\caption{\em Boxplots for the correlation coefficients between the estimated score and true treatment effect
with three different methods applied to survival outcomes. The empty and filled boxes represent low and high dimensional ($p=50$ and $p=1000$) cases, respectively.
Left upper panel: moderate main effect and independent covariates; right upper panel: moderate main effect and correlated covariates; left lower panel: big main effect and independent covariates; right lower panel: big main effect and correlated covariates.}
 \label{fig:surv}
\end{figure}

In top left, top right, bottom left and bottom right panels of Figure \ref{fig:surv}, we plotted the boxplots of the correlation coefficients between the estimated scores $\hat{\bgamma}'\bZ$ and $\Delta(\bZ)$ under  simulation settings, (1), (2), (3) and (4), respectively. The patterns were similar to those for the continuous and binary responses and confirmed our findings that the ``efficiency-augmented method'' performed the best among the three methods in general.

\section{Examples}
\label{sect:example}
 It has been known that the breast cancer can be classified into different subtypes using gene expression profile and the effective treatment may be different for different subtypes of the disease \citep{Loi:2007}. In this section, we apply the proposed method to study the potential interactions between gene expression levels and Tamoxifen treatment in the breast cancer patients.

 The data set consists of 414 patients in the cohort GSE6532 collected by \cite{Loi:2007} for the purpose of characterizing ER-positive subtypes with gene expression profiles. The dataset including demographic information and gene expression levels can be downloaded from the website {\it www.ncbi.nlm.nih.gov/geo/query/acc.cgi?acc=GSE6532}. Excluding patients with incomplete information, there are 268 and 125 patients receiving Tamoxifen and alternative treatments, respectively. In addition to the routine demographic information, we have $44,928$ gene expression measurements for each of the 393 patients. The outcome of the primary interest here is the distant metastasis free survival time, which subjects to right censoring due to incomplete follow-up. The metastasis free survival times in two treatment groups are not statistically different with a two-sided $p$ value of 0.59 based on the log-rank test (Figure \ref{example0}). The goal of the analysis is to construct a score using gene expression levels for identifying subgroup of patients who may or may not be benefited from the Tamoxifen treatment in terms of the distant metastasis free survival. To this end, we select the first 90 patients in the Tamxifen arm and an equal number of patients in the alternative treatment arm to form the training set and reserve the rest 213 patients as the independent validation set. In selecting the training and validation sets, we use the original order of the observations in the dataset without additional sorting to ensure an objective analysis.

 We first identify 5,000 genes with highest empirical variances and then construct an interaction score by fitting the Lasso penalized Cox regression model with modified covariates based on the 5,000 genes in the training set. The Lasso penalty parameter is selected via 20-fold cross-validation. The resulting interaction score is a linear combination of expression levels of seven genes. Here, a low interaction score favors Tamoxifen treatment. We apply the gene score to classify the patients in the validation set into high and low score groups according to if her score is greater than the median level. In the high score group, the distant metastasis free survival time in the Tamoxifen group is shorter than that in the alternative group with an estimated hazard ratio of 3.52 for Tamoxifen versus non-Tamoxifen treatment group (logrank test $p=0.064$). In the low score group, the distant metastasis free survival time in the Tamoxifen group is longer than that in the alternative group with an estimated hazard ratio of 0.694 ($p=0.421$). The estimated survival functions of both treatment groups are plotted in the upper panels of Figure {\ref{example1}.  The interaction between constructed gene score and treatment is statistically significant in the multivariate Cox regression based on the validation set ($p=0.004$).

  Furthermore, we implement the efficiency augmentation method and obtain a new score, which is based on expression level of eight genes. Again, we classify the patients in the validation set into high and low score groups based on the constructed gene score. In the high score group, the distant metastasis free survival time in the Tamoxifen group is shorter than that in the alternative group with a $p$ value of 0.158. The estimated hazard ratio is 2.29 for Tamoxifen versus non-Tamoxifen treatment group. In the low score group, the distant metastasis free survival time in the Tamoxifen group is longer than that in the alternative group with an estimated hazard ratio of 0.828. The $p$ value from the logrank test is not significant ($p=0.697$). The estimated survival functions of both treatment groups are plotted in the middle panels of Figure {\ref{example1}. The separation is slightly worse than that based the gene score constructed without augmentation.  The interaction between constructed gene score and treatment is also statistically significant at 0.05 level ($p=0.025$).

  For comparison purpose, we also fit a multivariate Cox regression model with treatment, the gene expression levels, and all treatment-gene interactions as the covariates. Lasso penalty is selected via 20-fold cross validation. The resulting gene score is a single gene based on the estimated treatment-gene interaction term of the Cox model. However, the interaction score fails to stratify the population according to the treatment effect in the validation set. The results are shown in the lower panel of Figure {\ref{example1}. The interaction between the constructed gene score and treatment is not statistically significant ($p=0.29$).

  To further objectively examine the performance of the proposal in this data set, we randomly split the data into training and validation sets and construct the score measuring individualized treatment effect in the training sets with three methods: ``new", ``new/augmented'' and ``full regression''. Patients in the test set are then stratified into high and low score groups. We calculate the difference in log hazard ratio for Tamoxifen versus non-Tamoxifen treatment between high and low score groups. A positive number indicates that women in low score group benefitted more from Tamoxifen treatment than those in high score group as the model indicates. In Figure {\ref{example2}}, we plot the boxplot of the differences in the log hazard ratio based on 100 random splitting. To speed the computation, all scores are constructed using only 2500 genes with top empirical variances. The results indicate that the proposed and the corresponding augmented methods tend to perform better than the common full regression method and this observation is consistent with our previous findings based on simulation studies.

As a limitation of this example, the treatment is not randomly assigned to the patients as in a standard randomized clinical trial. Therefore, the results need to be interpreted with caution. In addition, the sample size is limited and further verification of the constructed gene score with independent data sets is needed.

\begin{figure}[ht] \centering
\resizebox{!}{3.5 in}{\rotatebox{0}{\includegraphics{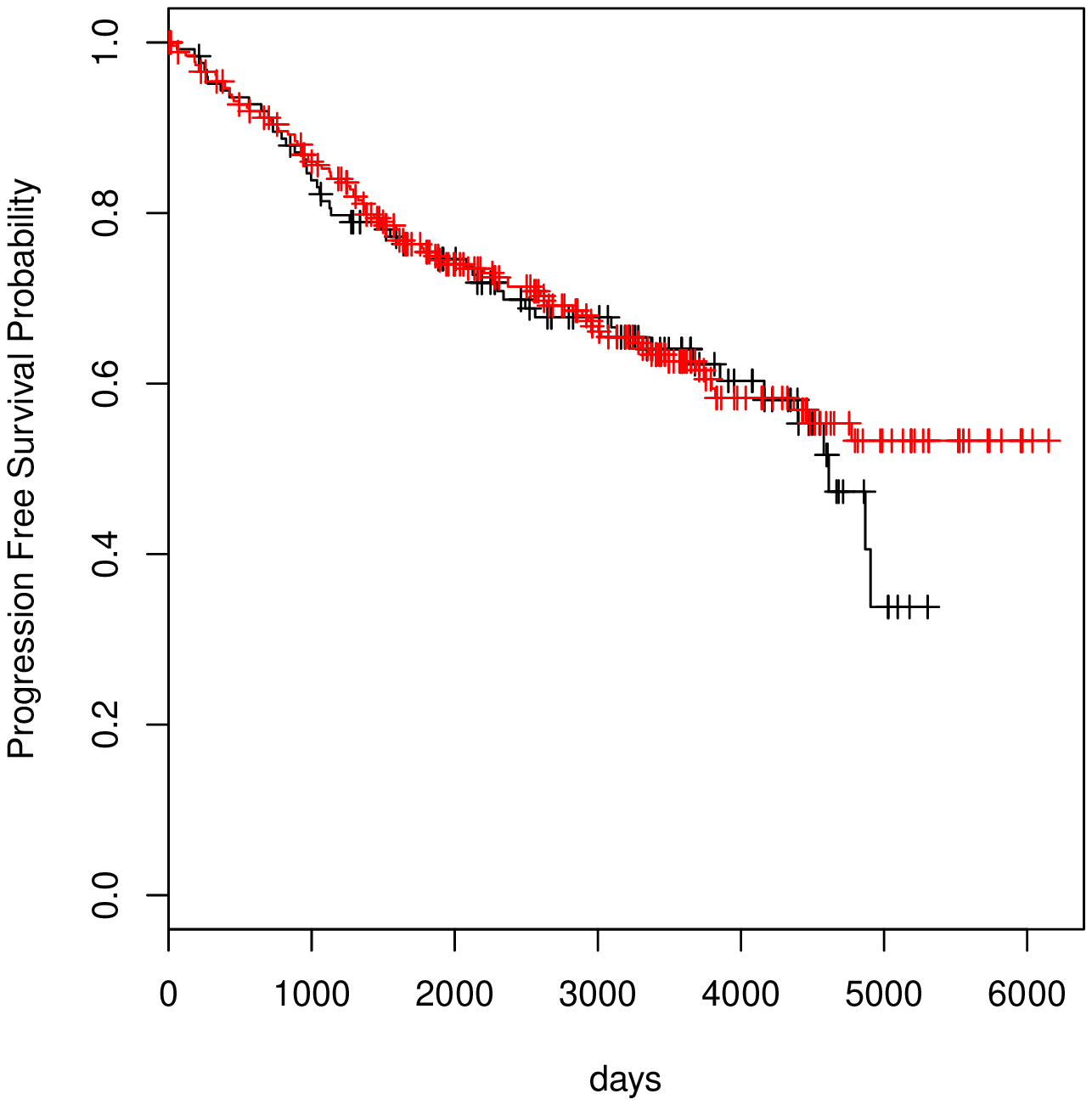}}}
\caption{Survival functions of the Tamoxifen and alternative treatment groups in 393 breast cancer patients. red line, Tamoxifen treatment group; black line, alternative treatment group } \label{example0}
\end{figure}

\begin{figure}[ht] \centering
\resizebox{!}{5.0in}{\rotatebox{0}{\includegraphics{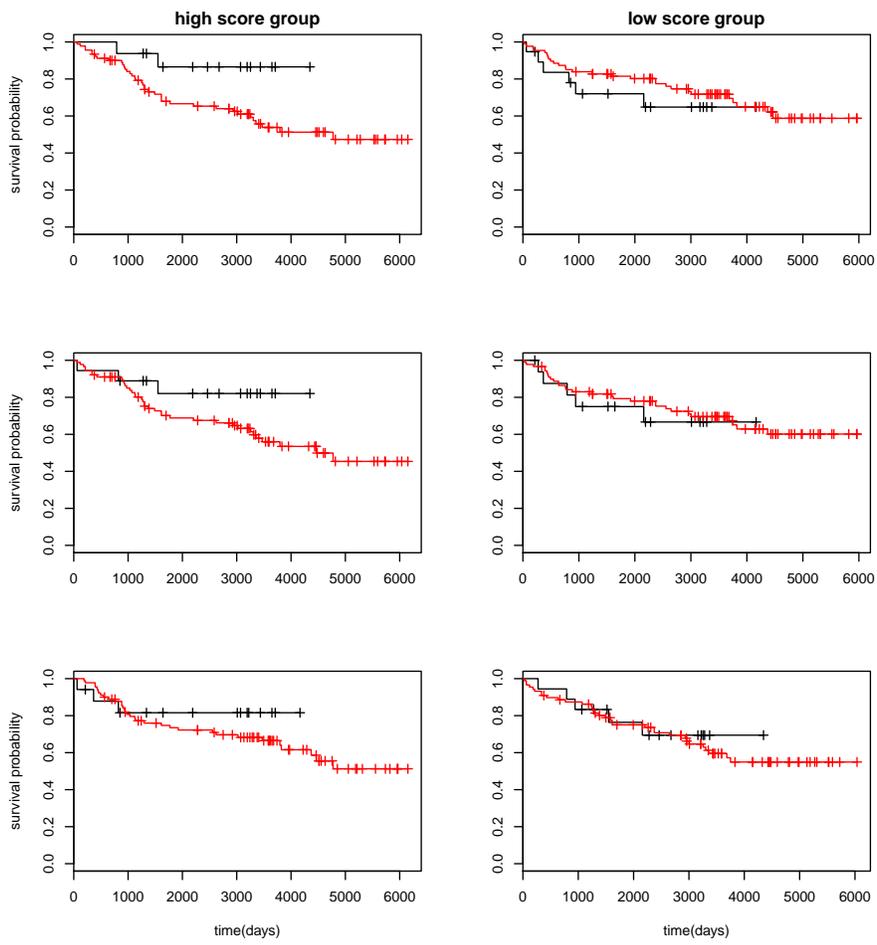}}}
\caption{Survival functions of the Tamoxifen and alternative treatment groups stratified by the interaction score in the test sets: red line, Tamoxifen treatment group; black line, alternative treatment group. Upper panels: the score based on the ``new'' method; middle panels: the score based on ``new/augmentated'' method; lower panel: the score is based on ``full regression'' method. } \label{example1} \end{figure}

\begin{figure}[ht] \centering
\resizebox{!}{3.5 in}{\rotatebox{0}{\includegraphics{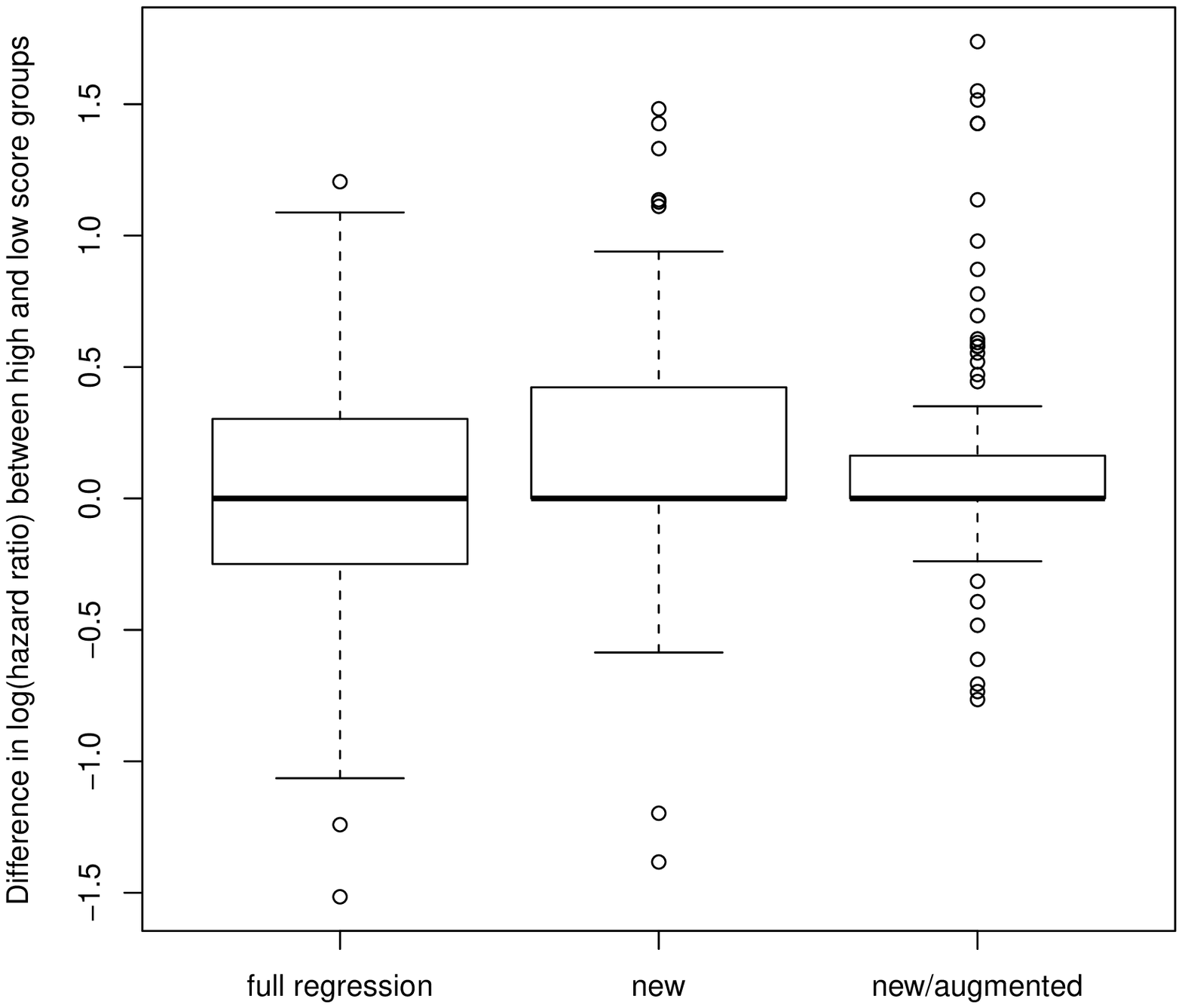}}}
\caption{Boxplots for differences in log(hazard ratio) between high and low risk groups based on 100 random splitting on GSE6532. The big positive number represents high quality of the constructed score in stratifying patients according to individualized treatment effect.} \label{example2} \end{figure}

\section{Discussion} In this paper we have proposed a simple method to explore the potential
interactions between treatment and a set of high dimensional covariates. The
general idea is to use $\bW(\bZ)\cdot T/2$ as  new covariates
in a regression or generalized regression  model to predict the outcome variable.
The resulting linear
combination $\hat{\bgamma}'\bW(\bZ)$ is then used to stratify  the patient
population. A simple efficiency augmentation procedure can be used to improve the performance of the method.

The proposed method can be used in much broader
way. For example, after creating  the modified covariates $\bW(\bZ)\cdot T/2$, other
data mining techniques such as PAM and support vector machines can also be used to link the new covariates with the outcomes \citep{Fr91, pam2003, hastie06:_discus_suppor_vector_machin_applic}. Most
dimension reduction methods in the literature can be readily adapted
to handle the potentially high dimensional covariates.  For univariate
analysis, we also may perform  large scale hypothesis testing on the
modified data, to  identify a list of covariates having interaction with the treatment; one could for example directly use the
Significance Analysis of Microarrays (SAM) method   \citep{SAM02} for this purpose.  Extensions in these
directions are promising and warrant further research.

Lastly,  the proposed method can also be used to analyze data from observational studies.
 However, the constructed interaction score may lose the corresponding causal interpretation. On the other hand, if a reasonable propensity score model is available, then we still can implement the modified covariate approach on matched or reweighted data such that the resulted score still retains
the appropriate causal interpretation \citep{Rosenbaum:Rubin:1983}.

\bibliographystyle{plainnat}
\bibliography{paperref}

\section{Appendix}
\subsection{Justification of the objective function based on the working model}

Under the linear working model for continuous response, we have
$$ \mbox{E} \{l(Y, f(\bZ)T)|\bZ, T=1\}= \frac{1}{2}\left[\mbox{E}\{ (Y^{(1)})^2|\bZ\}-2m_1(\bZ)f(\bZ)+f(\bZ)^2\right]$$
and
$$ \mbox{E} \{l(Y, f(\bZ)T)|\bZ, T=-1\}=\frac{1}{2}\left[\mbox{E}\{ (Y^{(-1)})^2|\bZ\}+2m_{-1}(\bZ)f(\bZ)+f(\bZ)^2\right],$$
where $m_t(\bz)=\mbox{E}(Y^{(t)}|\bZ=\bz)$ for $t=1$ and -1. Therefore
\begin{align*}
 {\cal L}(f)=&\mbox{E}\{l(Y, f(\bZ)T)\}\\
     =&\mbox{E}_{\bZ}\left[\frac{1}{2} \mbox{E}_Y\{l(Y, f(\bZ)T)|\bZ, T=1\}+\frac{1}{2} \mbox{E}_Y\{l(Y, f(\bZ)T)|\bZ, T=-1\} \right]\\
     =&\mbox{E}_{\bZ}\left(\left[\frac{1}{2}\{m_1(\bZ)-m_{-1}(\bZ)\}-f(\bZ)\right]^2 \right)+\mbox{constant}.
\end{align*}
Therefore, the minimizer of this objective function
$$f^*(\bz)=\frac{1}{2}\{m_1(\bz)-m_{-1}(\bz)\}=\frac{1}{2}\Delta(\bz)$$ for all $\bz \in \mbox{Support of } \bZ.$

$$ $$
Under the logistic working model for binary response, we have
$$ \mbox{E}\{l(Y, f(\bZ)T)|\bZ, T=1\}= m_1(\bZ)f(\bZ)-\log(1+e^{f(\bZ)}),$$
and
$$ \mbox{E}\{l(Y, f(\bZ)T)|\bZ, T=-1\}=-m_{-1}(\bZ)f(\bZ)-\log(1+e^{-f(\bZ)}).$$
Thus
\begin{align*}
 {\cal L}(f)=& \mbox{E}\{l(Y, f(\bZ)T)\}\\
            =&\mbox{E}_{\bZ}\left[\frac{1}{2} \mbox{E}_Y\{l(Y, f(\bZ)T)|\bZ, T=1\}+\frac{1}{2} \mbox{E}_Y\{l(Y, f(\bZ)T)|\bZ, T=-1\} \right]\\
            =&\frac{1}{2}\mbox{E}_{\bZ}\left[\Delta(\bZ)f(\bZ)-\log(1+e^{f(\bZ)})-\log(1+e^{-f(\bZ)}) \right].
\end{align*}
Therefore
$$\frac{\partial {\cal L}(f)}{\partial f}=\frac{1}{2} \mbox{E}_{\bZ}\left[\Delta(\bZ)-\frac{1-e^{f(\bZ)}}{1+e^{f(\bZ)}} \right],$$
which implies that the minimizer of ${\cal L}(f)$
$$f^*(\bz)=\log\frac{1-\Delta(\bz)}{1+\Delta(\bz)}$$
for all $\bz \in \mbox{Support of } \bZ$ or equivalently
$$\Delta(\bz)=\frac{1-e^{f^*(\bz)}}{1+e^{f^*(\bz)}}.$$

Alternatively, under the logistic working model with binary response, we may focus on the objective function
$$\tilde{l}(Y, f(\bZ)T)=(1-Y)f(\bZ)T-Ye^{-f(\bZ)T}.$$
Therefore
$$ \mbox{E}\{\tilde{l}(Y, f(\bZ)T)|\bZ, T=1\}= \{1-m_1(\bZ)\}f(\bZ)-m_1(\bZ)e^{-f(\bZ)},$$
and
$$ \mbox{E}\{\tilde{l}(Y, f(\bZ)T)|\bZ, T=-1\}=-\{1-m_{-1}(\bZ)\}f(\bZ)-m_{-1}(\bZ) e^{f(\bZ)}.$$
Thus
\begin{align*}
 {\cal L}(f)=& \mbox{E}\{\tilde{l}(Y, f(\bZ)T)\}\\
            =&\mbox{E}_{\bZ}\left[\frac{1}{2} \mbox{E}_Y\{l(Y, f(\bZ)T)|\bZ, T=1\}+\frac{1}{2} \mbox{E}_Y\{l(Y, f(\bZ)T)|\bZ, T=-1\} \right]\\
            =&\mbox{E}_{\bZ}\left[\frac{1}{2}\{m_1(\bZ)-m_{-1}(\bZ)\}f(\bZ)-\frac{1}{2}m_1(\bZ)e^{-f(\bZ)}-m_{-1}(\bZ) e^{f(\bZ)} \right]
\end{align*}
Therefore
$$\frac{\partial {\cal L}(f)}{\partial f}=\frac{1}{2} \mbox{E}_{\bZ}\left[\{m_1(\bZ)-m_{-1}(\bZ)\}+m_1(\bZ)e^{-f(\bZ)}-m_{-1}(\bZ) e^{f(\bZ)}  \right]$$
which implies that the minimizer of ${\cal L}(f)$
$$f^*(\bz)=\log\frac{m_{1}(\bz)}{m_{-1}(\bz)}$$
for all $\bz \in \mbox{Support of } \bZ.$

$$ $$
Under the Cox working model for survival outcome, we have
\begin{align*}
 \mbox{E}_Y\{l(Y, f(\bZ)T)|\bZ, T\}=& \mbox{E}_Y \left(\int_0^\tau \left[ Tf(\bZ)-\log\{\mbox{E}(e^{Tf(\bZ)}I(\tilde{X}\ge u))\}\right]d  N(u)|\bZ, T\right)\\
                                     =& \int_0^\tau \left[ f(\bZ)-\log\{\mbox{E}(e^{Tf(\bZ)}I(\tilde{X}\ge u))\}\right] \mbox{E} \left\{I(\tilde{X} \ge u)|\bZ, T\right\}\lambda_T(u; \bZ)du
 \end{align*}
where $\lambda_t(u; \bZ)$ is the hazard function for $\tilde{X}^{(t)}$ given $\bZ$ for $t=1/-1.$ Since
$$ {\cal L}(f)= \mbox{E}_{\bZ}\left[\frac{1}{2} \mbox{E}_Y\{l(Y, f(\bZ)T)|\bZ, T=1\}+\frac{1}{2} \mbox{E}_Y\{l(Y, f(\bZ)T)|\bZ, T=-1\} \right] $$
\begin{align*}
\frac{\partial {\cal L}(f)}{\partial f}=&\frac{1}{2}\mbox{E}\int_0^\tau \biggm \{ I(\tilde{X}^{(1)}\ge u)\lambda_1(u; \bZ)-I(\tilde{X}^{(-1)}\ge u)\lambda_{-1}(u; \bZ)\\
&-e^{f(\bZ)}I(\tilde{X}^{(1)}\ge u)\Lambda(u; f)+e^{-f(\bZ)}I(\tilde{X}^{(-1)}\ge u)\Lambda(u; f) \biggm \}du,
\end{align*}
 where $$\Lambda(t; f)=\frac{\mbox{E}[I(\tilde{X} \ge u)\{\lambda_1(u; \bZ)+\lambda_{-1}(u; \bZ)\}]}{\mbox{E}\{e^{Tf(\bZ)}I(\tilde{X}\ge u)\}}.$$
 Setting the derivative at zero, the minimizer $f^*(\bz)$ satisfies
 \begin{align*}
  &e^{f^*(\bz)}\mbox{E}\{\Lambda^*(\tilde{X}^{(1)})|\bZ=\bz\}-e^{-f^*(\bz)}\mbox{E}\{\Lambda^*(\tilde{X}^{(-1)})|\bZ=\bz\}\\
 =& \mbox{Prob}(C^{(1)}>X^{(1)}|\bZ=\bz)-\mbox{Prob}(C^{(-1)}>X^{(-1)}|\bZ=\bz)
 \end{align*}
for all $\bz \in \mbox{Support of } \bZ,$ where $\Lambda^*(u)=\Lambda(u, f^*).$ When censoring rates are the same in two arms for all given $\bz,$
$$ f^*(z)=-\frac{1}{2}\log \left[ \frac{E \{\Lambda^*(\tilde{X}^{(1)})|\bZ=\bz\}}{E\{\Lambda^*(\tilde{X}^{(-1)})|\bZ=\bz\}}\right]$$

\subsection{Justification of the optimal $a_0(\bz)$ in the efficient augmentation}
Let  $S(y, \bw^*, \bgamma)$ be the derivative of the objective function $l(y, \bgamma'\bw^*)$ with respect to $\bgamma.$ $\hat{\bgamma}$ is the root to an estimating equation
$$Q(\bgamma)=N^{-1}\sum_{i=1}^N S(Y_i, \bW_i^*, \bgamma)=0.$$
Similarly, the augmented estimator $\hat{\bgamma}_a$ can be viewed as the root of the estimating equation
$$Q_a(\bgamma)=N^{-1}\sum_{i=1}^N \left\{S(Y_i, \bW_i^*, \bgamma)-T_i\cdot \ba(\bZ_i)\right\}=0,$$
 Since $E\{T_i \cdot \ba(\bZ_i)\}=0$ due to randomization, the solution of the augmented estimating equation always converges to the $\bgamma^*$ in probability.
 It is straightforward to show that
 $$\hat{\bgamma}-\bgamma^*=N^{-1}A_0^{-1}\sum_{i=1}^N S(Y_i, \bW_i^*, \bgamma^*)+o_P(N^{-1})$$
 and
 $$\hat{\bgamma}_a-\bgamma^*=N^{-1}A_0^{-1}\sum_{i=1}^N \{S(Y_i, \bW_i^*, \bgamma^*)-T_i \ba(\bZ_i)\}+o_P(N^{-1})$$
 where $A_0$ is the derivative of $\mbox{E}\{S(Y_i, \bW_i^*, \bgamma)\}$ with respect to $\bgamma$ at $\bgamma=\bgamma^*.$ Selecting the optimal $\ba(\bz)$ is equivalent to
 minimizing the variance of $\{S(Y_i, \bW_i^*, \bgamma^*)-T_i \ba(\bZ_i)\}.$
 Noting that
 $$\mbox{E}\left[\{S(Y_i, \bW_i^*, \bgamma^*)-T_i \ba(\bZ_i)\}^{\otimes 2}\right]=\mbox{E}\left[\{S(Y_i, \bW_i^*, \bgamma^*)-T_i \ba_0(\bZ_i)\}^{\otimes 2}\right]+\mbox{E}[\{\ba(\bZ_i)-\ba_0(\bZ_i)\}^{\otimes 2}],$$
 where $\ba_0(\bz)$ satisfies the equation
 $$ E \left[\{S(Y, \bW^*, \bgamma^*)-T \ba_0(\bZ)\} T \eta(\bZ)\right]=0$$
 for any function $\eta(\cdot)$, $\ba_0(\cdot)$ is the optimal augmentation term minimizing the variance of $\hat{\bgamma}_a.$
 Since $\ba_0(\cdot)$ is the root of the equation
 $$ E\left[\{S(Y, \bW^*, \bgamma^*)-T \ba_0(\bZ)\}'T \biggm| \bZ \right]=0,$$
$$\ba_0(\bz)=\frac{1}{2}\left[\mbox{E}\{S(Y, \bW(\bz)/2, \bgamma^*)|\bZ=\bz, T=1\}- \mbox{E}\{S(Y, -\bW(\bz)/2, \bgamma^*)|\bZ=\bz, T=-1\}\right].$$
$$~~$$

For continuous response, $$S(Y, \bgamma'\bW^*)=-\frac{1}{2}T\bW(\bZ)\left\{Y-\frac{1}{2}T\bW(\bZ)'\bgamma \right\}$$ and
\begin{align*}
a_0(\bz)=&\frac{1}{2}\left(\mbox{E}[-\bW(\bz)\{Y-\bW(\bz)'\bgamma^*/2\}/2|T=1, \bZ=\bz]-\mbox{E}[\bW(\bz)\{Y+\bW(\bz)'\bgamma^*/2\}/2|T=-1, \bZ=\bz]\right)\\
=& -\bW(\bz)\left\{\frac{1}{4}\mbox{E}(Y|T=1, \bZ=\bz)+\frac{1}{4}\mbox{E}(Y|T=-1, \bZ=\bz)\right\}\\
=& -\frac{1}{2}\bW(\bz)\mbox{E}(Y|\bZ=\bz)
\end{align*}
$$~$$
For binary response,
$$S(Y, \bgamma'\bW^*)=-\frac{1}{2}\bW(\bZ)T \left\{Y-\frac{e^{T\bW(\bZ)'\bgamma/2}}{1+e^{T\bW(\bZ)'\bgamma/2}}\right\}$$ and
\begin{align*}
a_0(\bz)=& -\frac{1}{4}\bW(\bz)\left[\mbox{E} \left\{Y-\frac{e^{\bW(\bz)'\bgamma^*/2}}{1+e^{\bW(\bz)'\bgamma^*/2}}\biggm| T=1, \bZ=\bz\right\}+\mbox{E} \left\{Y-\frac{e^{-\bW(\bz)'\bgamma^*/2}}{1+e^{-\bW(\bz)'\bgamma^*/2}} \biggm| T=-1, \bZ=\bz\right\}\right]\\
=& -\frac{1}{4} \bW(\bz)\left\{\mbox{E}(Y|T=1, \bZ=\bz)+\mbox{E}(Y|T=-1, \bZ=\bz)-\left(\frac{e^{\bW(\bz)'\bgamma^*/2}}{1+e^{\bW(\bz)'\bgamma^*/2}}+\frac{e^{-\bW(\bz)'\bgamma^*/2}}{1+e^{-\bW(\bz)'\bgamma^*/2}} \right)\right\}\\
=& -\frac{1}{2}\bW(\bz)\left\{\mbox{E}(Y|\bZ=\bz)-\frac{1}{2}\right\}
\end{align*}
$$~$$
For survival response, the estimating equation based on the partial likelihood function is asymptotically equivalent to the estimating equation $N^{-1}\sum_{i=1}^N S(Y_i, \bW^*_i, \bgamma)=0,$ where
$$S(Y, \bW^*, \bgamma)=-\int_0^\tau \left[\bW^*-\mathbf{R}(u; \bgamma^*)\right] M(du, \bW^*, \bgamma^*).$$
Thus, $$\ba_0(\bz)=-\frac{1}{2}\left[\frac{1}{2}\bW(\bz)\left\{G_1(\tau; \bz)+G_2(\tau; \bz)\right\}-\int_0^\tau \mathbf{R}(u) \{G_1(du; \bz)-G_2(du; \bz)\}\right],$$

\subsection{Lasso algorithm in the efficient augmentation}
In general, the augmentation term is in the form of
$a_0(\bZ_i)=\bW(\bZ_i)'\hat{r}(\bZ_i),$
where $\hat{r}(\bZ_i)$ is a simple scalar. The lasso regularized objective function can be written as
$$\frac{1}{N}\sum_{i=1}^N \left\{ l(Y_i, \bgamma' \bW_i^*)-\bgamma' \bW_i^*\hat{r}(\bZ_i)\right\}+\lambda |\bgamma|.$$
In general, this lasso problem can be solved iteratively. For example, when $l(\cdot)$ is the log-likelihood function of the logistic regression model, then
with we may update $\hat{\bgamma}$ iteratively by solving the standard OLS-lasso problem
$$\frac{1}{N}\sum_{i=1}^N \hat{w}_i(\hat{z}_i-\bgamma'\bW_i^*)^2 +\lambda|\bgamma|$$
where  $\hat{\bgamma}$ is the current estimator for $\bgamma,$
$$\hat{z}_i=\hat{\bgamma}'\bW_i^*+\hat{w}_i^{-1}\{Y_i-\hat{p}_i-\hat{r}(\bZ_i)\}, ~~~~~\hat{w}_i=\hat{p}_i(1-\hat{p}_i)$$
and
$$\hat{p}_i=\frac{\exp\{\bgamma'\bW_i^*\}}{1+\exp\{\bgamma'\bW_i^*\}}.$$

\end{document}